\pdfoutput=1 
\documentclass[reprint,superscriptaddress,aps,pra,floatfix]{revtex4-2} 

\usepackage{amsmath}  
\usepackage{graphicx}

\usepackage{hyperref}
\usepackage{orcidlink}

\usepackage{float}    
\usepackage{multirow}
\usepackage{makecell}

\begin{document}

\title{A Covalent Bonding Network Using \texorpdfstring{$p$}{p}-block Elements to Stabilize Lithium-Rich Cathode Materials}
\author{Amelia Zhao}
\affiliation{Department of Materials Science and Engineering, Stanford University, Stanford, CA}
\affiliation{Stanford Institute for Materials and Energy Sciences, SLAC National Accelerator Laboratory, Menlo Park, CA}
\author{Edward Mu}
\affiliation{Department of Chemistry, Stanford University, Stanford, CA}
\author{Jessica Luu}
\affiliation{Department of Chemistry, Stanford University, Stanford, CA}
\author{John Vinson \orcidlink{0000-0002-7619-7060}}
\affiliation{Material Measurement Laboratory, National Institute of Standards and Technology, Gaithersburg, MD}
\author{Brian Moritz}
\affiliation{Stanford Institute for Materials and Energy Sciences, SLAC National Accelerator Laboratory, Menlo Park, CA}
\author{Eder G. Lomeli}
\email{ederlomeli@lbl.gov}
\affiliation{Advanced Light Source, Lawrence Berkeley National Laboratory, Berkeley, CA}
\author{Thomas P. Devereaux}
\email{tpd@stanford.edu}
\affiliation{Department of Materials Science and Engineering, Stanford University, Stanford, CA}
\affiliation{Stanford Institute for Materials and Energy Sciences, SLAC National Accelerator Laboratory, Menlo Park, CA}
\affiliation{Geballe Laboratory for Advanced Materials, Stanford University, Stanford, CA}

\date{\today}

\begin{abstract}
Li-rich battery cathodes (Li$_x$M$_y$O$_z$, where $x>y$) have sparked significant attention for their high energy densities, but their practical application is currently limited by structural instabilities associated with high-valent redox. Recent work has demonstrated reversible high-potential (de)lithiation with low voltage hysteresis in Li$_4$FeSbO$_6$ (LFSO), but the stabilization mechanism for this unique charging behavior is not fully understood, limiting our ability to apply insights from LFSO to more earth-abundant cathodes. In this paper, we present a covalent Sb-O bonding network in LFSO which is facilitated by the $p$-block orbital character of Sb. Using \textit{ab initio} computational modeling, we identify this bonding state and indicate its contributions to crystal stability; we also identify a signature for this bonding network using theoretical and experimental element-specific X-ray spectroscopy. Substituting Sb with other $p$-block elements preserves this stabilizing bond, pointing to a family of new cathode chemistries beyond LFSO. This work establishes a design principle for development of sustainable energy-dense batteries: $p$-block elements like Sb possess specific orbital-mediated oxygen bonding which contributes to stability of high-valent layered oxide cathodes. 


\end{abstract}
\maketitle

\section{Introduction}

Lithium-ion batteries have powered the consumer electronics and electric vehicle industries for over four decades, but remain insufficiently cost-effective and energy-dense to satisfy the critical need for grid-scale energy storage. One limiting factor is the performance of the cathode, where conventional materials face trade-offs between capacity and stability, and require expensive, rare elements such as cobalt \cite{yuanyuan_research_2025, ziegler_determinants_2021, goodenough_li-ion_2013}.

A proposed mechanism to address these limitations is using lithium-rich layered oxide cathodes: materials whose Li content exceeds standard transition metal (TM) layer stoichiometry (\textit{i.e.}, Li$_x$M$_y$O$_z$, where $x>y$). Due to the increased number of added/removed Li per formula unit, Li-rich cathodes offer significant improvements in specific capacity and energy density compared to conventional cathodes, and rely on more earth-abundant TMs like Ni, Mn, and Fe \cite{he_challenges_2021, manthiram_nickel-rich_2016, arifiadi_toward_2026}. However, Li-rich cathodes suffer from instability and voltage hysteresis \cite{mccalla_understanding_2015, croy_examining_2013, assat_fundamental_2018, zheng_li-_2017}, rendering them impractical for commercial applications. 

These structural and performance issues, especially at high voltages, have been associated with the formation of oxygen dimers and gas release, triggering irreversible capacity loss~\cite{assat_editors_2016, xu_elucidating_2018, mccalla_visualization_2015, hu_evolution_2018, ning_inhibition_2020, chen_unraveling_2019, sharifi-asl_oxygen_2019}. 
As voltage increases, the standard picture of electrochemistry in these materials 
has been that TM (cationic) redox gives way to oxygen (anionic) redox, allowing continued Li extraction at the cost of increased ligand hole density and deleterious effects on battery performance~\cite{gent_design_2020, assat_fundamental_2018, assat_fundamental_2017}. Recent work has challenged this framework, using first-principles simulation and spectroscopic measurements to demonstrate that high-valent redox actually occurs due to negative charge transfer and hole formation in delocalized TM-O hybridized states~\cite{ramachandran_formal_2026, hsu_transition-metal-oxygen_2026, lomeli_negative_2025}. 
Ultimately, the development of practical Li-rich cathodes will require formalized design rules for high-voltage stability, which are informed by these novel insights into charge-transfer redox. These rules should then be implemented toward novel material chemistries, optimizing stable energy density with more earth-abundant elemental compositions.

In this work, we identify non-redox-active covalent bonding states as potential drivers of stability in Li-rich cathodes, specifically Li$_4$FeSbO$_6$ (LFSO). LFSO exhibits high-valent cycling, accessing negative-charge-transfer Fe-O states to allow for reversible removal of up to two Li$^+$ per formula unit \cite{ramachandran_formal_2026}. We connect this unusual stability to the $p$-block orbital properties and bonding structure of Sb. Using density functional theory (DFT) and theoretical bond evaluation techniques, we identify a covalent state between Sb and O in both lithiated and delithiated LFSO, which does not exist when Sb is replaced with a $d$-block 5$^+$ element. Based on thermodynamic, structural, and electronic analyses, we propose that this covalent bonding network is a key factor for the reversible cycling stability of LFSO. Finally, we search across a series of theoretical cathode materials with Sb replaced by chemically similar, but more earth-abundant elements, and identify the same stabilizing bonding state in these novel materials. Our results provide a bonding-focused interpretation for electrochemical cycling behavior and offer a pathway toward more stable, high-valent Li-rich cathodes through tuning orbital interactions. 

\section{Results and Discussion}

\subsection{An orbital-mediated Sb-O bonding network in LFSO}

\begin{figure} [h!]
    \centering
    \includegraphics[width=\linewidth]{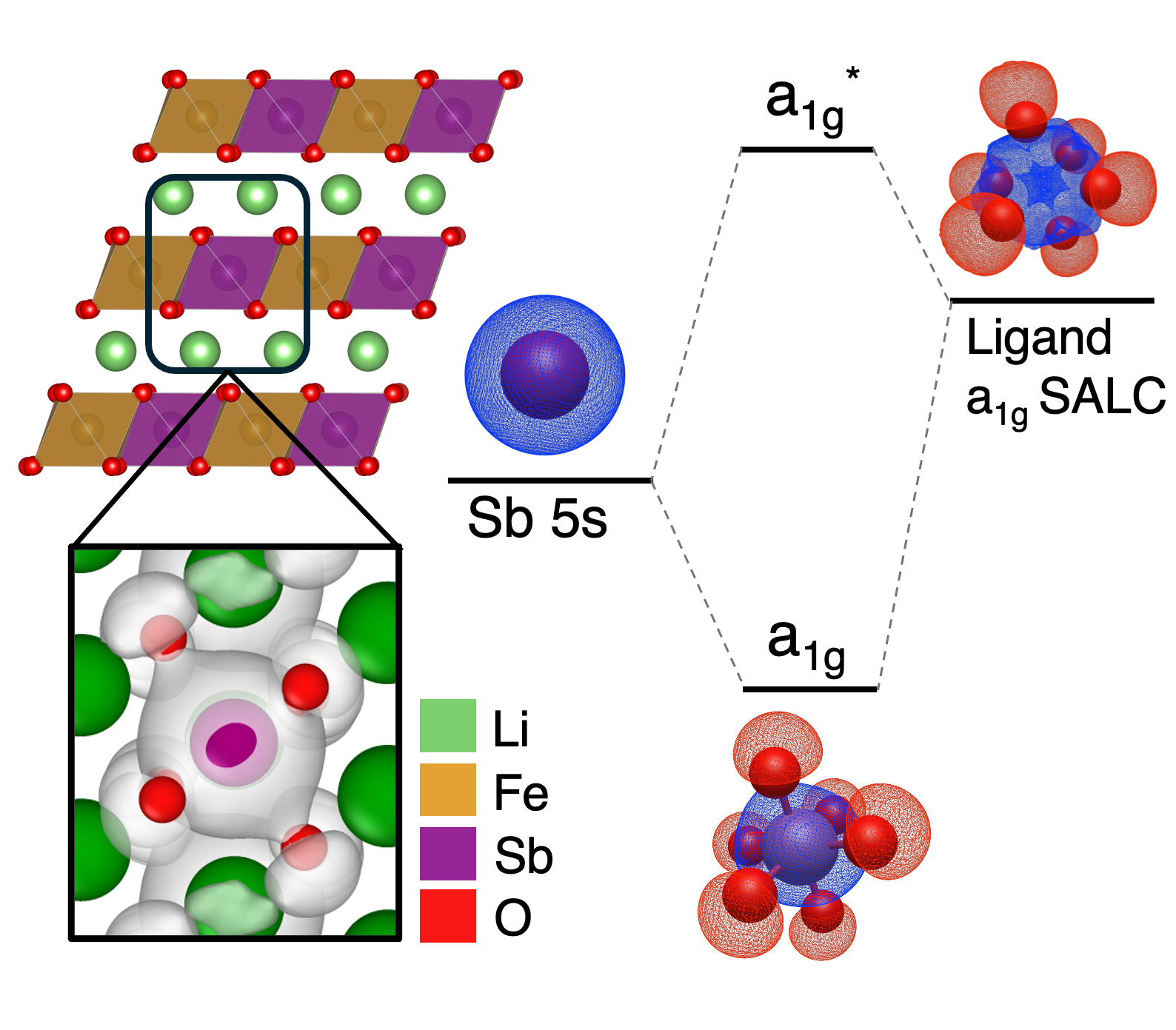}
    \caption[] {\textbf{A bonding network in the TM layer of LFSO.} LFSO is a lithium-rich layered oxide cathode material with $O3$ stacking order in its pristine state. Here, we focus on the $a_{1g}$ molecular orbital (shown in the inset, lower right), formed by covalent overlap between Sb and the symmetry adapted linear combination (SALC) of its octahedrally coordinated O $2p$ atomic orbitals.}
    \label{fig:overview}
\end{figure}

We focus on the role of Sb and its covalency with its octahedral oxide ligands to understand the high-valent stability of LFSO. Unlike other redox-inactive transition metals typically used in Li-rich cathodes, such as Mn, Ti, and Zr \cite{kim_electrochemical_2004, lu_lack_2003, thackeray_li2mno3-stabilized_2007, yu_crystalline_2016, worku_li-rich_2025}, Sb sits in the $p$-block of the periodic table. Its $5s$-orbital contributions are energetically comparable with the ligand symmetry-adapted linear combination (SALC) composed of the in-phase combination of the $2p$ orbitals of its octahedrally coordinated oxygens. Thus, the Sb-O $a_{1g}$ molecular bonding orbitals have significant contribution from both metal orbitals and ligand group orbitals, rather than being mostly ligand $p$ character as we would expect from $d$-block TM-centered octahedra. These bonding orbitals between Sb and its six coordinated oxygens form a highly-covalent bond network throughout the metal layer. The same Sb $5s$-O $2p$ hybridization also produces a higher-energy antibonding $a_{1g}^*$ state, demonstrating how this interaction redistributes oxygen orbital character between strongly bonding and antibonding combinations (Fig. \ref{fig:overview}). 

\begin{figure*}[!htb]
\centering
  \includegraphics[width=\textwidth]{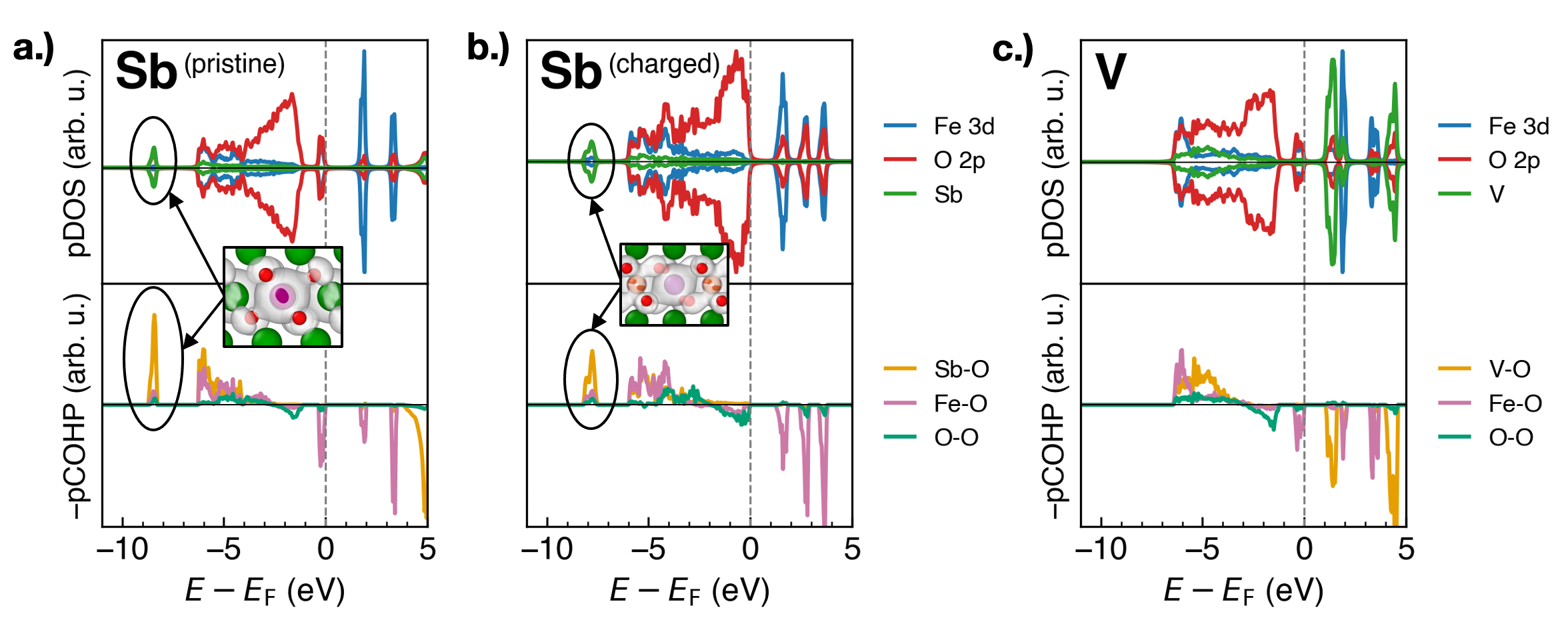}
  \caption[Identification of bonding state via LFSO vs. LFVO DOS comparison.]{\textbf{Identification of $p$-block bonding state via LFSO vs. LFVO DOS comparison.} (a) Element- and orbital-resolved pDOS and pCOHP for pristine LFSO. The bonding state is highlighted, with a visualization of the real-space charge density associated with the highlighted Sb-O electronic band. In the lower panel, moving upward along the y-axis indicates a more bonding character for the given atomic interaction, while moving downward indicates a more antibonding interaction. (b) pDOS and pCOHP for delithiated LFSO, with a persistent Sb-O covalent bond. (c) pDOS and pCOHP for an analogous non-$p$-block structure, with V replacing Sb. We find no comparable V-O bonding state. }
  \label{fig:lfso_pdos}
\end{figure*}

We use DFT to calculate the element- and orbital-projected density of states (pDOS) for pristine LFSO (Supp. Fig. \ref{fig:ball_and_stick_LFSO}) to identify this molecular bonding state (Fig. \ref{fig:lfso_pdos}a). The pDOS shows an occupied state deep below the Fermi level which consists almost entirely of Sb and O 2$p$ orbital contributions. This indicates a covalent bond between the two which is characteristic of the $p$-block orbital configuration and ligand interactions of Sb. Details for DFT and all other calculations are provided in Supp. Info. Section 1, Methods. 

The exact nature of this state is confirmed using bonding analysis and real-space visualization. With the {\sc lobster} electronic structure processing package, we conduct projected crystal orbital Hamilton population (pCOHP) analysis on our pDOS \cite{dronskowski_pcohp_1993}. This calculation partitions our energy range into antibonding (positive pCOHP), nonbonding, and bonding (negative pCOHP) regions, and projects these regions onto specific atom pair interactions. The sharp negative COHP peak at the energy corresponding to the Sb-O electronic state indicates that this interaction is a strongly covalent bond. Projected charge density for the electronic bands that make up this feature is visualized, depicting a spherical $s$-orbital contribution from Sb with significant overlap with the adjacent in-phase lobes of O-$2p$ orbitals (Fig. \ref{fig:lfso_pdos}a). These results further confirm our understanding of our identified state: it is strongly covalent, bonding, and consists largely of contributions from Sb and O. 

Similar pDOS and pCOHP analysis demonstrates that this state also exists in delithiated LFSO, with two Li$^+$ per formula unit removed (Fig. \ref{fig:lfso_pdos}b). Previous X-ray diffraction (XRD) measurements and multiplet calculations have shown that upon delithiation, LFSO undergoes both an $O$3-$O$1 structural phase transition \cite{delmas_structural_1980} and an electronic transition to a negative-charge-transfer ground state with significant oxygen hole density \cite{ramachandran_formal_2026}. Despite this, the Sb-O covalent network maintains strong bonding character in our pDOS and pCOHP. Additionally, we do not observe a notable increase in O-O covalent bonding states in our pCOHP, with the only O-O pCOHP contributions coming from weak oxygen interactions across different octahedra. Visualization of the partial charge density of the Sb-O bonding network in the delithiated structure shows that the shape of the octahedral Sb-O molecular orbital remains largely constant between lithiated and delithiated LFSO, with more electron density on Fe in the charged structure. 

A comparison with the pDOS and pCOHP of a Li-rich cathode material which does not contain a $p$-block metal shows that this state results from an Sb-specific interaction. This new material is constructed using the same layered oxide structure as LFSO ($O$3 stacking and honeycomb in-plane ordering); however, we replace Sb with V and geometrically optimize to generate Li$_4$FeVO$_6$ (LFVO). Vanadium was chosen because Sb$^{5+}$ and V$^{5+}$ have the same formal oxidation number, but sit in the $p$-block vs. the $d$-block of the periodic table, respectively, with valence electron configurations [Kr]$4d^{10}5s^25p^3$ for Sb vs. [Ar]$3d^34s^2$ for V. Choosing two closed-shell cations with the same formal oxidation state controls for the effects of spin polarization and frontier-shell occupancy differences. At the same time, the distinction between the orbital energies and symmetries of Sb vs. V should strongly impact each metal's capacity to form covalent bonds with octahedrally coordinated ligands. In particular, the $s$-orbital energy levels of V are very different from O $2p$ orbitals, meaning there should be very little V contribution to the octahedral $a_{1g}$ molecular bonding state. 

As expected, no V-O covalently bonded state (Fig. \ref{fig:lfso_pdos}c) is found down to energies of 23 eV below the Fermi level, well below where one would expect the lowest-energy interacting orbitals of O. Instead, the majority of the density of states of V are antibonding $d$-orbital contributions in the conduction band. Consequently, we conclude that the Sb-O covalent bonding feature in LFSO stems from its orbital configuration and electronic properties as a Group 15, or pnictogen, atom with a filled $d$ shell. 

\subsection{Discussion of bonding network and stability}

After identifying this covalent bonding network in LFSO, we then aim to understand its contributions toward cathode stability. The first method for this is comparison with the convex hull phase diagram. We use DFT at the Strongly Constrained and Appropriately Normed (SCAN) \cite{sun_strongly_2015} level of theory to calculate energies for all known stable structures in the compositional phase space of LFSO and LFVO, a critical step to ensure that our analysis is accurate while using meta-generalized gradient approximation (GGA) functionals. We find that pristine LFSO is on the convex hull with an energy 22 meV/atom below any known combination of reference structures among the 45 stable compounds in its phase space given by the Materials Project database \cite{jain_commentary_2013}, indicating that this structure is stable against decomposition and likely to be synthesizable. In comparison, LFVO is 64 meV/atom above the convex hull, exceeding the $\approx$50 meV/atom threshold typically used as a cutoff for crystal metastability. Charged LFSO is 40 meV/atom above the convex hull, meaning that this structure is thermodynamically metastable and remains more stable than LFVO despite the changes in stacking order, electronic structure, and stoichiometry during charging. This can also be compared with a similar structure, charged LFVO, which is constructed by replacing Sb with V in the charged LFSO structure, and which lies 81 meV/atom above the convex hull. From these results, we conclude that the stability of LFSO is related to the specific properties of Sb, and in particular, its bonding network with O. 

To further confirm this, we investigate the role of Sb in the structural changes that occur during LFSO delithiation. Unlike the Fe-O bond distances in our simulated cell, which shrink significantly during delithiation due to negative-charge-transfer covalency and orbital contraction, we find minimal change in the Sb-O bond length between the pristine and charged structures. These results are verified by Sb and Fe \textit{K}-edge extended X-ray absorption fine structure (EXAFS) measurements. Between the spectra collected on pristine and charged (4.3 V) LFSO, Fe shows both a significant edge shift and real-space Fe-O nearest-neighbor peak shift (Supp. Fig. \ref{fig:fe_exafs}), consistent with the computational results and previous experimental findings \cite{ramachandran_formal_2026}. In contrast, Sb shows negligible edge shift and no change to the Sb-O bond length (Supp. Fig. \ref{fig:exafs}). Therefore, the local Sb environment is preserved upon delithiation, which we associate with the redox-inactive Sb-O covalent bonding state. Furthermore, the simulated O-O distance does not greatly contract upon charging, going from an average of 2.861 \AA{} to 2.748 \AA{}. This aligns with our pCOHP bonding/antibonding analysis (Fig. \ref{fig:lfso_pdos}b) where we see minimal evidence of O-O bonding states, and signifies a lack of oxygen dimerization, which typically occurs between 1.2~\AA{} and 1.6 \AA{} in Li-rich cathode materials \cite{chen_unraveling_2019, mccalla_visualization_2015}. This analysis aids our understanding of the charging reversibility of LFSO, since O-O dimer species are well-known to introduce voltage decay and cycling instability to electrode materials \cite{ning_inhibition_2020, chen_lithium_2016, sharifi-asl_oxygen_2019}. We suggest that covalency between Sb and its coordinated oxygens contributes to LFSO's high-voltage stability by reducing the physical proximity of Fe octahedra, even with the additional structural pressures introduced by negative charge transfer. All bond distances, calculated using fractional atomic positions with periodic boundary conditions, can be found in Supp. Table \ref{table:bonds}. 

Finally, we examine how covalent bonding impacts the electronic stability of the crystal. Previous work on Sb doping in Mn-rich Na-ion cathodes proposed that Sb introduces electron density to its neighboring O, which prevents O lattice migration \cite{jamil_sb-doped_2024}. Additionally, studies on Na-ion cathodes with other $p$-block dopants, such as Sn and Zn, concluded that metal-oxygen covalent bonding induces charge delocalization on oxygen which suppresses irreversible redox reactions \cite{li_tuning_2026}, and that this charge redistribution is attributed to the electron attractiveness of $4s$ orbitals compared to $3d$ orbitals \cite{zou_unveiling_2025}. We believe that these observations arise from the effects of a covalent bonding network between metal $s$ orbitals and oxygen $p$ orbitals, analogous to the $5s-2p$ bonding interaction we have identified for Sb and O in LFSO. Bader analysis is consistent with this hypothesis, as we observe a depletion of charge from Sb between the pristine and delithiated LFSO structures, and a smaller charge difference if Sb is replaced with V in the same geometry-relaxed structures (Supp. Table \ref{table:bader}). We interpret that charge is being transferred into the covalent network, where Sb-O bonds act as a stabilization channel. Although this Sb-O bonding state remains occupied, hybridization and polarization of the shared oxygen orbitals may facilitate the redistribution of ligand hole density generated by Fe-O redox.

Together, these results indicate that the covalent bonding between Sb and O acts as a structurally and electronically stabilizing network throughout the TM layer. This supports previous experimental work, where Sb doping in Na and Li oxide cathodes has been shown to improve capacity retention, increase ion diffusion, and suppress undesirable phase transitions \cite{jamil_sb-doped_2024, cui_preparation_2011, ma_suppressing_2022}. These properties have been explained using the effects of Sb on lattice parameters and unit cell volume \cite{ma_suppressing_2022, cui_preparation_2011}. However, the unique orbital character and bonding properties of Sb are also critical for understanding the electrochemical behaviors of these cathodes, as well as LFSO. 

\subsection{Experimental characterization of the covalent bonding network }

After identifying this covalent bonding state as an important contributor to high-valent stability, we experimentally describe it using spectroscopic measurements and simulations. O $K$-edge resonant inelastic X-ray scattering (RIXS) is used to probe electronic de-excitation into our deep bonding state. We model RIXS measurements using the first-principles Bethe-Salpeter equation (BSE)-based {\sc ocean} code \cite{vinson_advances_2022,vinson_bethe-salpeter_2011}. To probe our bonding state, we take RIXS linecuts at three incident X-ray energies, corresponding to the three pre-edge peaks in our O $K$-edge X-ray absorption spectroscopy (XAS). Theoretical and experimental XAS spectra are provided in Supp. Fig. \ref{fig:xas}. As evidenced by the pDOS (Fig. \ref{fig:lfso_pdos}a), these three incident energies are associated with excitation into two Fe-O hole bands and one higher-energy Sb-O hole band. Here, this Sb-O hole band is the antibonding $a_{1g}^*$ combination of out-of-phase Sb 5s and O 2p orbitals (Fig. \ref{fig:overview}); that is, the antibonding counterpart to our identified covalent bonding network.

\begin{figure} [h!]
    \centering
    \includegraphics[width=\linewidth]{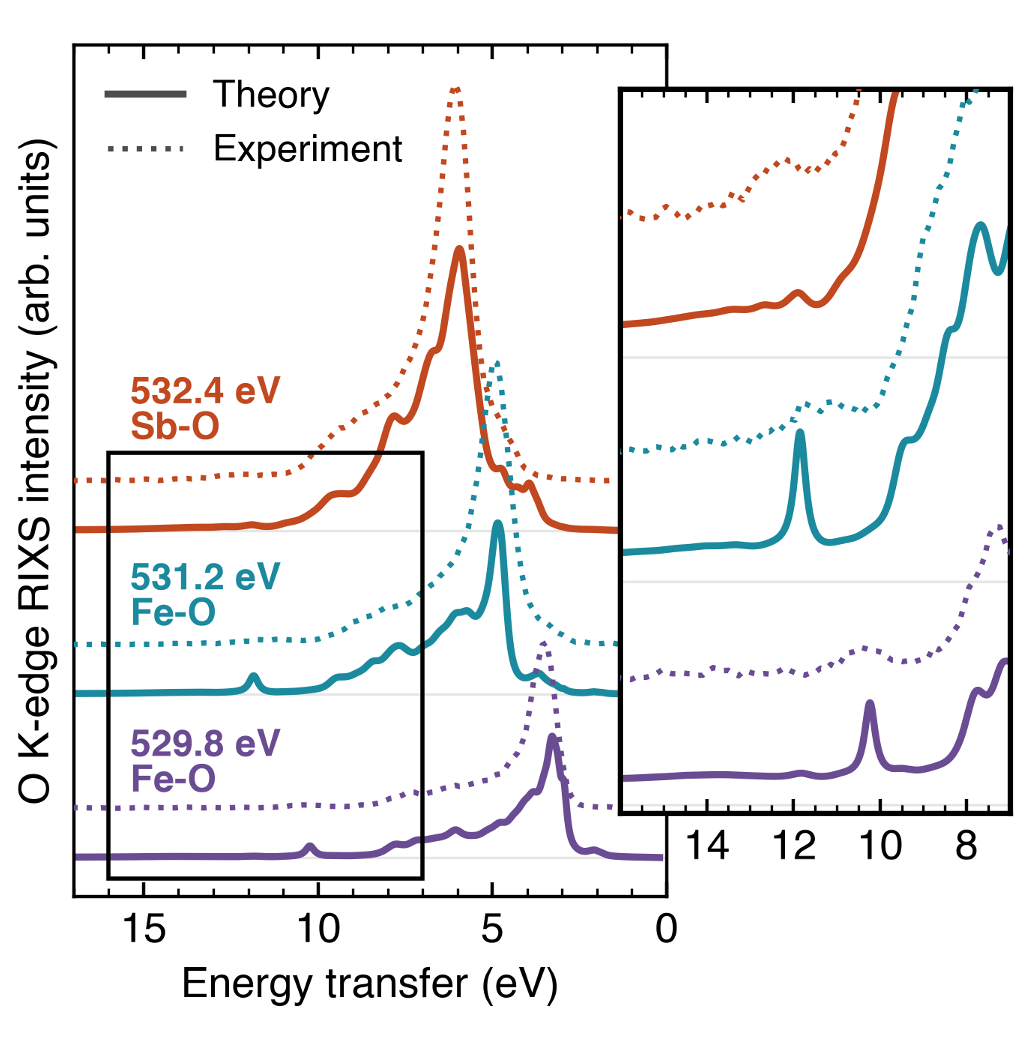}
    \caption[] {\textbf{Experimental signature for the bonding network in LFSO. } Experimental \cite{ramachandran_formal_2026} and theoretical oxygen $K$-edge RIXS linecuts for pristine LFSO, highlighting the feature that corresponds with de-excitation from the Sb-O covalent bonding network. }
    \label{fig:lfso_ocean_rixs}
\end{figure}

In both the experimental and theoretical RIXS linecuts, we see a distinct feature whose energy transfer is between 6.15~eV and 7 eV higher than the main emission line of occupied Fe-O bands near the Fermi energy. As this aligns well with the energetic positioning of our bonding state with respect to the Fermi energy in our calculated DOS (Fig. \ref{fig:lfso_pdos}a), we associate this feature with de-excitation from our covalent Sb-O bonding state. 

The experimental feature is suppressed compared to our theoretical prediction, which we attribute to broadening lifetime effects which grow with energy loss and are not included in our RIXS simulation. Additionally, the feature is sharpest for incident energies corresponding with excitation into unoccupied Fe-O antibonding states, compared to the energy exciting into the Sb-O antibonding state. This is due to the spatially extended and dispersive nature of the Sb-O $5s$-$2p$ hole band, which should broaden the RIXS feature, as well as interaction effects between the electron in the Sb-O antibonding state and the hole in the Sb-O bonding state which do not exist when exciting into an Fe-O antibonding state. The observation that the Sb-O bonding feature can be accessed through excitation into Fe-O-derived intermediate states is also consistent with the shared oxygen orbital character of the two octahedral environments, highlighting the electronic connectivity between the redox-active Fe-O states and the redox-inactive Sb-O bonding network. 

Ultimately, our RIXS analysis provides a spectroscopic signature for the stabilizing Sb-O covalent bonding network. This can be used to experimentally confirm the presence of this state and can act as a screening metric for potential new bonding-stabilized cathode materials.

\subsection{Materials search for stable, earth-abundant Li-rich cathode candidates}

This bonding network in LFSO provides a design rule toward informed selection of novel cathode material candidates. While LFSO demonstrates promising high-valent redox capabilities, it remains unsuitable for large-scale application due to the scarcity and cost of Sb. Thus, we aim to propose analogous Li-rich cathode structures that provide a stabilizing covalent network, but do not contain Sb. These materials may also exhibit high-valent reversibility and stable negative-charge-transfer ground states, due to the retained covalent bonding between the Sb replacement element and its oxygen ligands. 

When selecting replacement elements for Sb, our aim is to reproduce as many of its properties as possible. In particular, it is essential to choose $p$-block elements which have the valence orbital character to form covalent $a_{1g}$ bonds with their ligands. Thus, candidates in close proximity to Sb on the periodic are the most promising: elements in the pnictogen family (As, Bi) and neighboring elements in Group 14 (Ge, Sn, Pb). Although we predict that Group 16 elements (Se, Te, Po) should also exhibit a stabilizing bonding network, these are not our focus due to their scarcity and hazardous nature. To produce our theoretical material structures, Sb is directly replaced in LFSO with a substitute elemental candidate, and the new crystal is geometrically optimized. When using non-pnictogen elements, we charge balance our structure by replacing the Li in the TM layer with Mg, to avoid undesired charge density migration from Fe. The resulting candidate materials maintain the same stacking and in-plane structure as LFSO, allowing us to perform comparable analyses to investigate their bonding properties. 

\begin{figure*}[!htb]
\centering
  \includegraphics[width=0.9\textwidth]{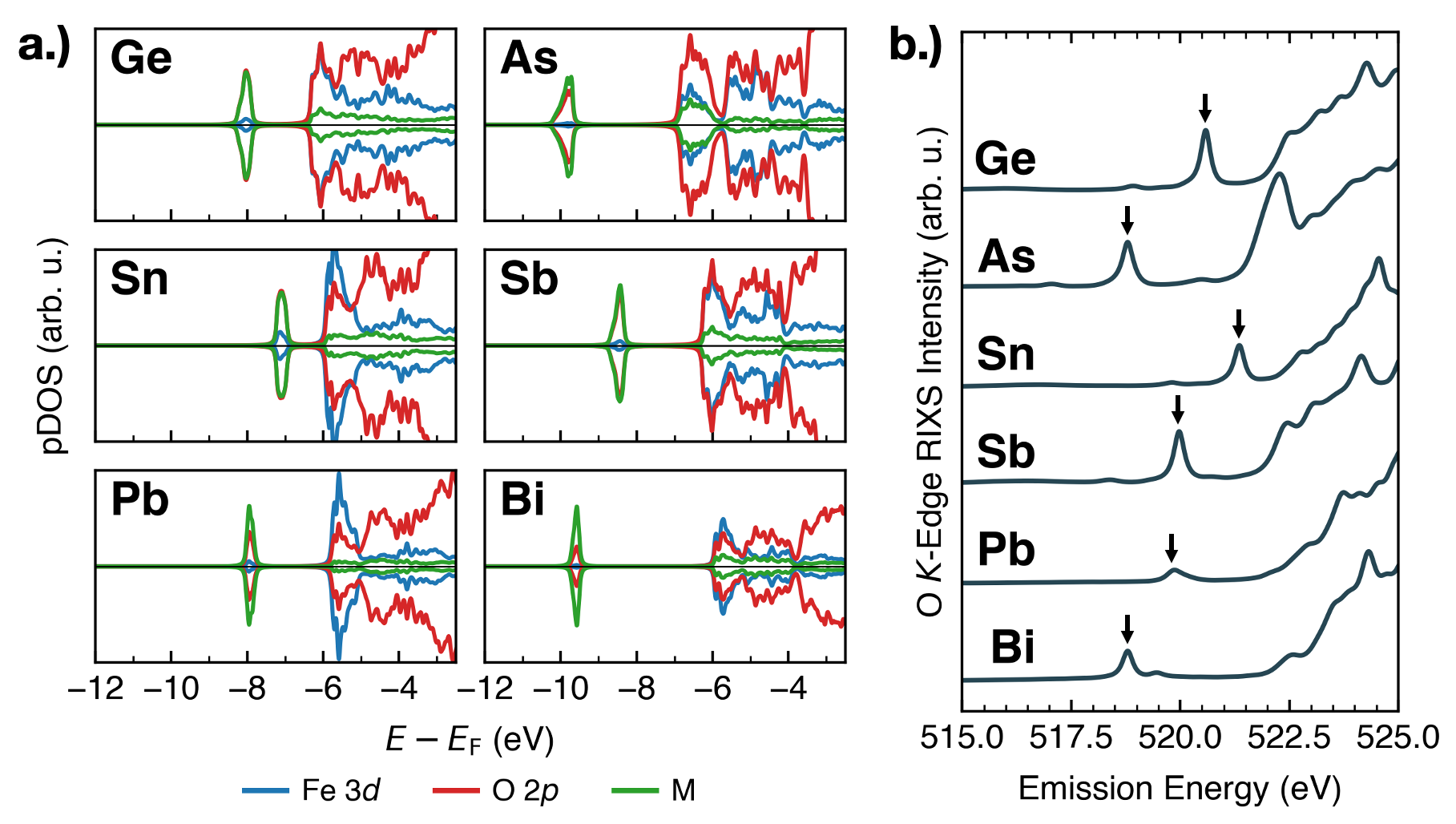}
  \caption[Electronic and spectroscopic characterization of covalent bonding in LFSO and Sb-replaced Li-rich cathode materials.]{\textbf{Electronic and spectroscopic characterization of covalent bonding in LFSO and Sb-replaced Li-rich cathode materials. } (a) pDOS zoom-in fragments for alternative cathode candidates showing the orbital-mediated bonding state. Individual plots are labeled with the replacement atom for Sb in a LFSO-like structure. (b) Simulated RIXS signature of the bonding state in Sb-replaced Li-rich cathodes. All spectra are calculated using {\sc ocean} at the O $K$-edge. Incident energies for each curve are chosen by taking the lowest-energy Fe-O peak of that structure's simulated XAS spectra.}
  \label{fig:six_panel_pdos_p}
\end{figure*}

pDOS analysis illustrates the presence of the covalent bonding network in each of these cathode material candidates (Fig. \ref{fig:six_panel_pdos_p}a). All of the new candidate materials exhibit the covalent state between the Sb replacement element and the $2p$ orbitals of its oxygen ligands, at an energy well below the `redox-active' iron-oxygen manifold near the Fermi level. This state contains nearly equal contributions from both metal (Sb/Sb replacement) and ligand, with the exception of Li$_4$FeBiO$_6$ and Li$_3$MgFePbO$_6$, which both have greater contribution from the metal $s$ orbital due to the larger atomic radii of Bi and Pb compared with Sb. These results are confirmed by pCOHP analysis and partial charge density visualization, which show that this state is, in fact, a covalent bonding state, and that the shape of the molecular orbital formed by the Sb-replacement octahedra closely resembles that of Sb in LFSO (Supp. Fig. \ref{fig:pdos_pcohp_6}). 

Using the {\sc ocean} spectroscopic modeling code, we confirm the experimental signature for this bonding network in each of our candidate cathode materials. For all Sb-replaced cathodes, we calculate O $K$-edge XAS spectra (Supp. Fig. \ref{fig:xas_6}) and RIXS linecuts (Fig. \ref{fig:six_panel_pdos_p}b). Like in LFSO, we observe a feature at high energy transfer (approximately 7~eV to 10 eV above the emission line), which we associate with our stabilizing bonding state. Figure~\ref{fig:six_panel_pdos_p}b shows only RIXS linecuts at incident energies corresponding with the first unoccupied Fe-O band, as defined by comparing theoretical XAS spectra with the calculated pDOS (Supp. Figs. \ref{fig:pdos_pcohp_6} and \ref{fig:xas_6}). As previously discussed for LFSO, the simulated RIXS feature for our bonding network is strongest at resonances which excite into the neighboring Fe-O octahedra, as opposed to (Sb-replacement)-O antibonding states. Regardless, the RIXS feature is still seen when exciting into any unoccupied state in all of our cathode candidate materials (Supp. Fig. \ref{fig:rixs_6}), confirming the presence of our bonding state even in materials without Sb and establishing RIXS as a suitable experimental technique for characterizing this bond network.

Finally, convex hull thermodynamic analysis gives preliminary insight into the generalized effect of this bonding state on crystal stability (Supp. Fig. \ref{fig:thermo_convex_hull} and Supp. Table \ref{table:convex_hull}). While none of these cathodes are as thermodynamically stable as LFSO, they are all more stable than LFVO and all lie within $\approx$50 meV/atom above the convex hull, classifying them as metastable. In particular, we find that Li$_3$MgFeGeO$_6$ and Li$_3$MgFeSnO$_6$ have unoccupied pDOS structures which are very similar to LFSO, and both lie $<5$ meV/atom above the convex hull, indicating that these structures are very nearly thermodynamically stable and are, in practice, not likely to decompose into more stable components. We recognize that the convex hull is an imperfect metric for synthesizability due to the impacts of varying reference structures and finite-temperature effects, as well as the fact that these calculations should be sensitive to the physical bond lengths of the Sb-replaced octahedra, which use the LFSO crystal as an initial structure \cite{bartel_review_2022, sun_thermodynamic_2016}. Nonetheless, these results suggest a connection between redox-inactive covalent bonding between the Sb replacement and its oxygen ligands, and the crystal's cycling stability. Thus, we propose that these are strong candidates as next-generation cathode materials, where the presence of a covalent bonding network contributes to crystal synthesizability and high-valent charging reversibility. 

\section{Conclusion}

In this work, we present an orbital-mediated Sb-O covalent bonding network in LFSO, which contributes to its unique high-voltage reversibility. We characterize this state using electronic density of states analysis and identify its experimental signature using spectroscopic measurements and first-principles simulation techniques. Treating the bonding network as a guiding principle for stability, we survey similar crystals which contain this bond, several of which substitute Sb with more earth-abundant elements. LFSO is a promising Fe-based cathode material due to its high operating potential and resistance to calendar aging. However, a mechanistic understanding of its high-valent stability is necessary to fully exploit these properties and translate them into new earth-abundant and scalable cathode materials. Our results address this, highlighting the importance of chemical bonding and orbital character as design parameters when selecting desirable cathode chemistries. 

This work contributes to a novel understanding of cathode electrochemical behavior through the lens of covalency and charge transfer. This framework is supported by RIXS theory and measurements, which allow us to identify shared electronic states and analyze orbital-specific correlations between occupied and unoccupied bands. In this picture for LFSO, redox happens during charging via negative charge transfer states in Fe-O octahedra. This process is simultaneously facilitated by redox-inactive bonding states in neighboring Sb-O octahedra, which are highly covalent and sit well below the Fermi energy. These bonding states act as scaffolding to ensure the crystal's structural integrity and as a charge channel to maintain electronic stability in the oxidized lattice. Ultimately, we propose that the covalent bonding of the non-redox-active elements, which stems from their electronic and orbital characteristics, are a key mechanism for the strong performance of LFSO. The strength of chemical bonding should be further explored as a path toward stable high-valent negative charge transfer redox in next-generation Li-rich cathodes. 

\section{Acknowledgments}
The work at Stanford University and SLAC National Accelerator Laboratory was supported by the U.S. Department of Energy, Office of Basic Energy Sciences, Division of Materials Sciences and Engineering, award number DE-SCL0000172. A.Z. acknowledges support from the National Science Foundation Graduate Research Fellowship Program under Grant No. DGE-2146755. Any opinions, findings, and conclusions or recommendations expressed in this material are those of the author(s) and do not necessarily reflect the views of the National Science Foundation. Calculations were performed on resources of the National Energy Research Scientific Computing Center (NERSC), a Department of Energy Office of Science User Facility, using NERSC award ERCAP0034968. 

NIST work was funded solely by the U.S. Department of Commerce. Certain software is identified in this paper in order to specify the experimental procedure adequately. Such identification is not intended to imply recommendation or endorsement of any product or service by NIST, nor is it intended to imply that the materials or equipment identified are necessarily the best available for the purpose.

This research used resources of the Advanced Light Source, which is a DOE, Office of Science User Facility, under contract number DE-AC02-05CH11231. Use of the Stanford Synchrotron Radiation Lightsource (SSRL), SLAC National Accelerator Laboratory, is supported by the U.S. Department of Energy, Office of Science, Office of Basic Energy Sciences under Contract No. DE-AC02-76SF00515.

The authors thank Joshua Kas and Jason Hwong for helpful discussions and advice. The authors also acknowledge Nic Liang for conducting Sb $K$-edge EXAFS measurements. 

\subsection{Author Contributions}
E.G.L. and T.P.D. conceived the study. A.Z. performed the {\sc vasp} DFT calculations and data analysis under the supervision of E.G.L., B.M., and T.P.D. A.Z. performed the {\sc ocean} XAS and RIXS simulation under the supervision of J.V. and E.G.L. E.M. and J.L. advised on chemical principles behind cathode candidate selection, and provided experimental structures and spectroscopic data. A.Z. wrote the first draft of the manuscript and all authors reviewed and approved the final version. 

\subsection{Competing Interests}
The authors declare no competing interests. 

\subsection{Data Availability}
Data is available from the authors upon reasonable request. 

\clearpage 

\bibliographystyle{apsrev4-2} 
\bibliography{refs} 

\clearpage
\onecolumngrid
\raggedbottom 

\begin{center}
  {\large\textbf{Supplementary Information}}\\[6pt]
  \textbf{A Covalent Bonding Network Using \texorpdfstring{$p$}{p}-block Elements to Stabilize Lithium-Rich Cathode Materials}\\[6pt]
  Amelia Zhao, Edward Mu, Jessica Luu, John Vinson, Brian Moritz, Eder G. Lomeli, Thomas P. Devereaux
\end{center}

\setcounter{section}{0}
\setcounter{figure}{0}
\setcounter{table}{0}
\renewcommand{\thesection}{S\arabic{section}}
\renewcommand{\thefigure}{S\arabic{figure}}
\renewcommand{\thetable}{S\arabic{table}}
\renewcommand{\theHsection}{S\arabic{section}}
\renewcommand{\theHfigure}{S\arabic{figure}}
\renewcommand{\theHtable}{S\arabic{table}}
\section{Methods}
\subsection{{\sc VASP} DFT calculations}

We used the projector-augmented wave (PAW) method in the Vienna \textit{ab initio} simulation package ({\sc vasp}) 6.4.3 to perform spin-polarized DFT geometric optimization, partial charge density visualization, and Bader analysis \cite{kresse_efficient_1996, kresse_efficiency_1996, kresse1999ultrasoft}. We used the Strongly Constrained and Appropriately Normed (SCAN) meta-generalized gradient approximation (meta-GGA) functional \cite{sun_strongly_2015}, including non-spherical contributions to the gradient of the density in PAW spheres. Our plane-wave energy cutoff was 520 eV for all calculations. Partial occupancies were determined using Gaussian smearing with width 0.2 eV for geometric optimization, and the tetrahedron method with Blöchl corrections without smearing \cite{blochl_improved_1994} was used for the single-point density-of-states calculations and bonding/antibonding analysis. We used a gamma-centered k-point grid of $6\times4\times6$ for all lithiated structures and $4\times6\times6$ for all delithiated structures. 

Pristine and delithiated LFSO structures were taken from experimental data and previous energetic optimization calculations \cite{ramachandran_formal_2026}. All other theoretical cathode materials were generated by manually replacing Sb with alternative elements, and replacing the Li$^+$ in the transition metal layer with Mg$^{2+}$ if necessary for charge balancing. We choose Mg$^{2+}$ based on its earth abundance and similar covalent radius to Li$^+$, and because crystals constructed with Mg$^{2+}$ have a lower energy above their convex hulls than analogous crystals constructed with Ca$^{2+}$ or Sr$^{2+}$. All structures are geometrically optimized to a force convergence of 0.01 eV/\AA{}. We used antiferromagnetic high-spin Fe in all structures, based on previous analysis which found that this was energetically favorable \cite{ramachandran_formal_2026}. 

\subsubsection{Convex hull analysis}
Thermodynamic convex hull phase diagrams for each structure were manually constructed using all stable entries in each respective chemical system from the Materials Project \cite{jain_commentary_2013, horton_accelerated_2025}. We relaxed each of these entries, including both ferromagnetic and antiferromagnetic configurations for Fe when relevant, to a force convergence of 0.02 eV/\AA{}. The SCAN functional was used for all calculations to ensure consistent energy comparisons. 

\subsubsection{Molecular orbital visualization}

Visualizations of molecular orbitals were generated using the {\sc orca} 6.1.1 quantum chemistry code \cite{ORCA6}. Molecular orbital structures were taken from previous {\sc vasp} DFT geometric optimizations. We used the SCAN level of theory, the def2-TZVP basis set \cite{weigend_balanced_2005}, and the Resolution of Identity (RI)-J approximation for Coulomb integrals with the def2/J auxiliary basis set \cite{weigend_accurate_2006}. 

\subsection{Bonding/antibonding analysis}

We used the Local-Orbital Basis Suite Towards Electronic-Structure Reconstruction (\textsc{lobster}) 5.1.1 package \cite{maintz_analytic_2013, maintz_lobster_2016, nelson_lobster_2020} for projected crystal Hamilton population (pCOHP), integrated COHP (iCOHP), and bond-distance analysis \cite{dronskowski_pcohp_1993, deringer_crystal_2011}. Inputs were generated using single-point calculations from {\sc VASP}-relaxed geometries, with the number of bands set to 140 to ensure a greater number of bands compared to atomic orbitals for all analyzed structures. We used the pbeVASPFit2015 atomic orbital basis set and calculated the COHP for all atomic interactions between 0.5~\AA{} and 3.0 \AA{}, including neighboring unit cells. Charge spilling for all tested structures was below 1.27$\%$, with all electrons recovered by projection. 

\subsection{BSE spectroscopy simulation}
We used the first-principles {\sc ocean} 3.2.0 spectroscopy simulation package \cite{vinson_bethe-salpeter_2011, gilmore_efficient_2015, vinson_advances_2022} to calculate theoretical X-ray absorption spectroscopy (XAS) and resonant inelastic X-ray scattering (RIXS) spectra within the Bethe-Salpeter equation (BSE) approximation. We modeled the oxygen {\it K}-edge for all calculations with incoming and outgoing photon polarizations summed over the $x$, $y$, and $z$ directions to reflect experimental conditions using powder samples.

Converged electron densities and wavefunctions from {\sc Quantum Espresso} were used as inputs for {\sc ocean}. We used the relaxed structures from {\sc vasp}, with the same k-point grids used for {\sc vasp} calculations, a plane-wave energy cutoff of 118 Ryd, and the same SCAN exchange-correlation functional approximation. We created pseudopotentials using the PseudoDojo library \cite{van_setten_pseudodojo_2018} and the Optimized Norm-Conserving Vanderbilt Pseudopotentials ({\sc oncvpsp}) program \cite{hamann_optimized_2013}. 

Based on a comparison with experimental XAS measurements for pristine and charged LFSO, we added an energy shift to our theoretical spectra to align their main edges. Emission energies for all structures are also calculated using a nominal edge energy shift taken from comparison with experimental RIXS measurements for pristine LFSO \cite{ramachandran_formal_2026}. 

\section{Supplementary Figures}

\begin{figure}[H]
    \centering
    \includegraphics[width=\textwidth]{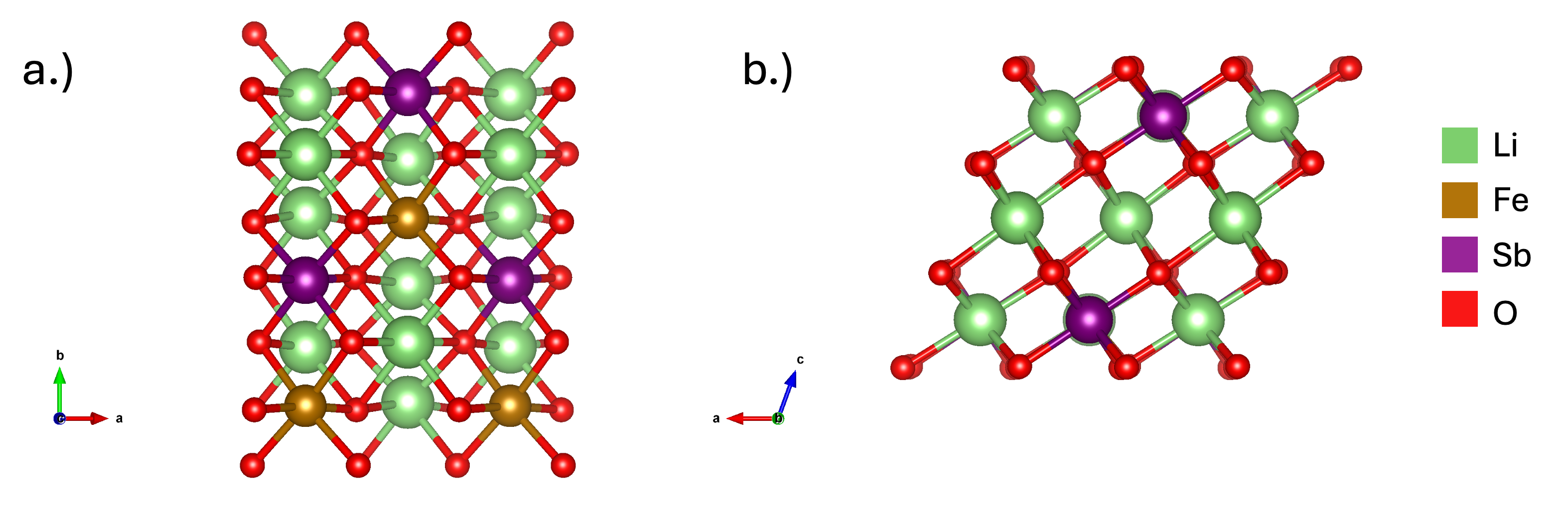}
    \medskip 
    \includegraphics[width=\textwidth]{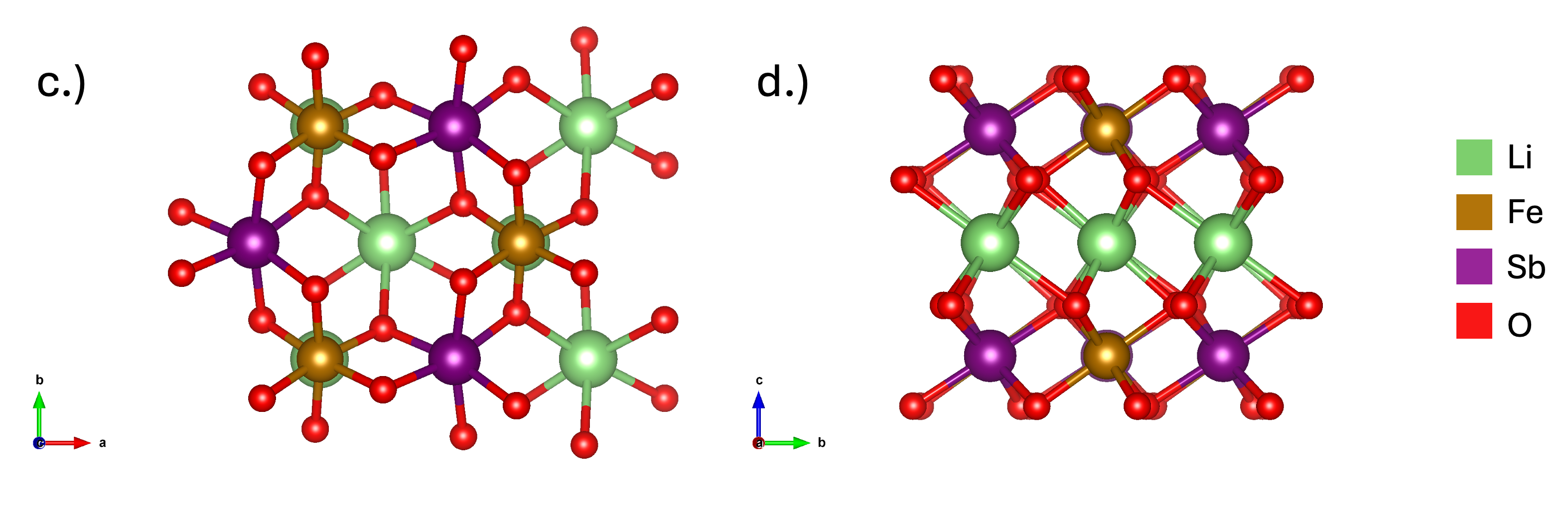}
    \caption{\textbf{Ball-and-stick models of pristine and charged LFSO used for DFT calculations.} Top and side views, respectively, are provided for pristine LFSO (a, b) and delithiated LFSO (c, d). Images generated using {\sc vesta} \cite{momma_vesta_2011}.}
    \label{fig:ball_and_stick_LFSO}
\end{figure}

\begin{figure}[H]
\centering
  \includegraphics[width=\textwidth]{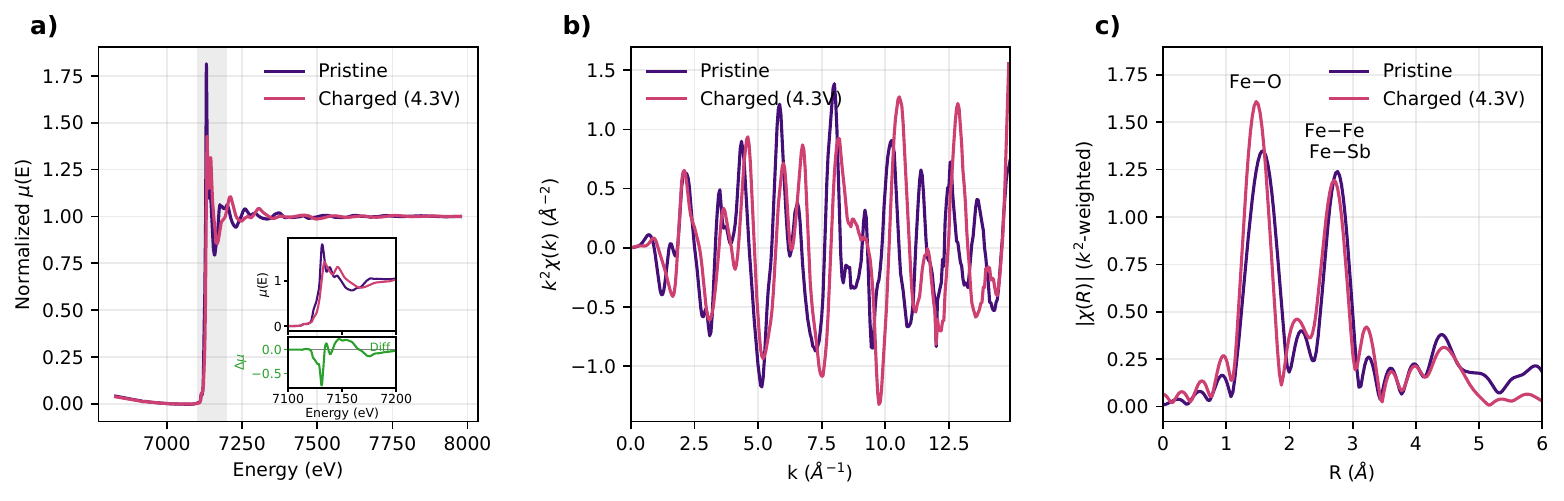}
  \caption[]{\textbf{Fe $K$-edge EXAFS measurements for pristine and charged LFSO.} (a) Experimental Fe $K$-edge X-ray absorption near-edge spectra (XANES) for pristine and charged LFSO, collected at the Stanford Synchrotron Radiation Lightsource (SSRL) beamline 4-1. A significant edge shift change is observed between the charged and pristine spectra (inset). (b) Fe $K$-edge extended X-ray absorption fine structure (EXAFS) data for pristine and charged LFSO, and (c) the magnitude of $\chi(R)$. Decreases in the first shell scatterers (Fe-O) and second-shell (Fe-Fe and Fe-Sb) are consistent with reduced bond distances upon charging observed by X-ray diffraction.}
  \label{fig:fe_exafs}
\end{figure}

\begin{figure}[H]
\centering
  \includegraphics[width=\textwidth]{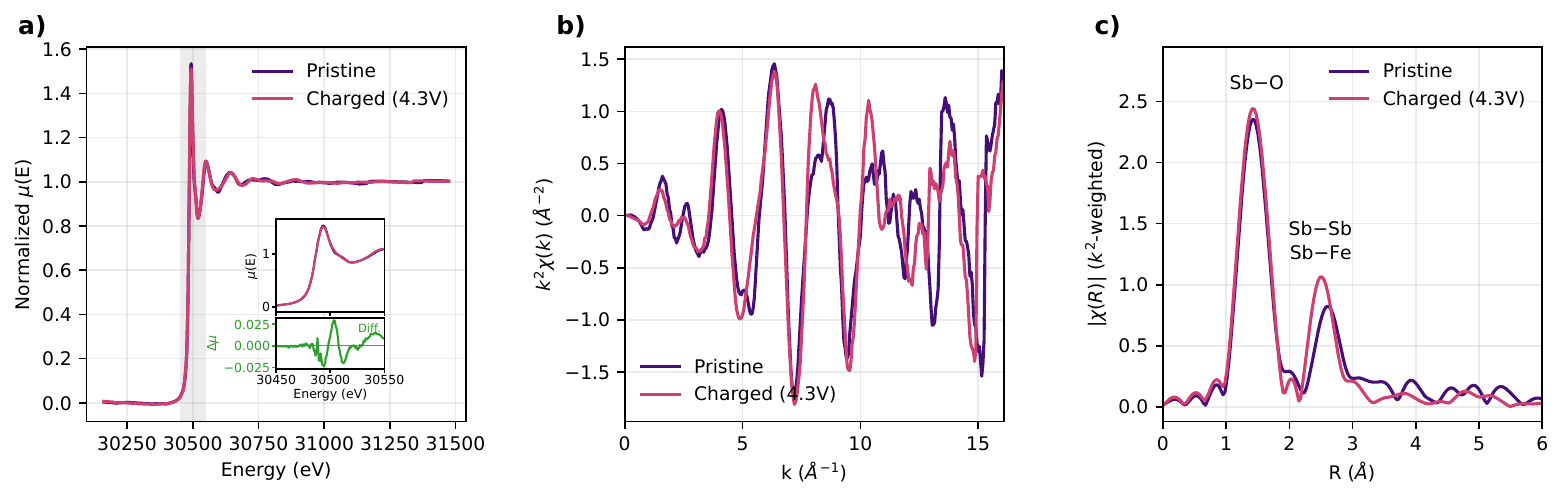}
  \caption[]{\textbf{Sb $K$-edge EXAFS measurements for pristine and charged LFSO.} (a) Experimental Sb $K$-edge X-ray absorption near-edge spectra (XANES) for pristine and charged LFSO, collected at SSRL beamline 4-1. No significant edge shift change is observed between the charged and pristine spectra, suggesting a retention of the Sb electronic structure (inset). (b) Sb K-edge extended X-ray absorption fine structure (EXAFS) data for pristine and charged LFSO, and (c) the magnitude of $\chi(R)$. The first coordination shell of Sb, that is, its bond distances with its octahedral nearest-neighbor oxygens (Sb-O scatterers), shows very little change upon charging to 4.3 V. This method measures instantaneous bond distances which are typically shorter than those measured by diffraction; however, the trends across the two methods remain consistent. The second-coordination shell, corresponding to Fe-Sb and Sb-Sb scatterers, decreases during charging, consistent with a reduction in Fe-O bond distances during the charging process.}
  \label{fig:exafs}
\end{figure}

\begin{figure}[H]
\centering
  \includegraphics[width=0.9\textwidth]{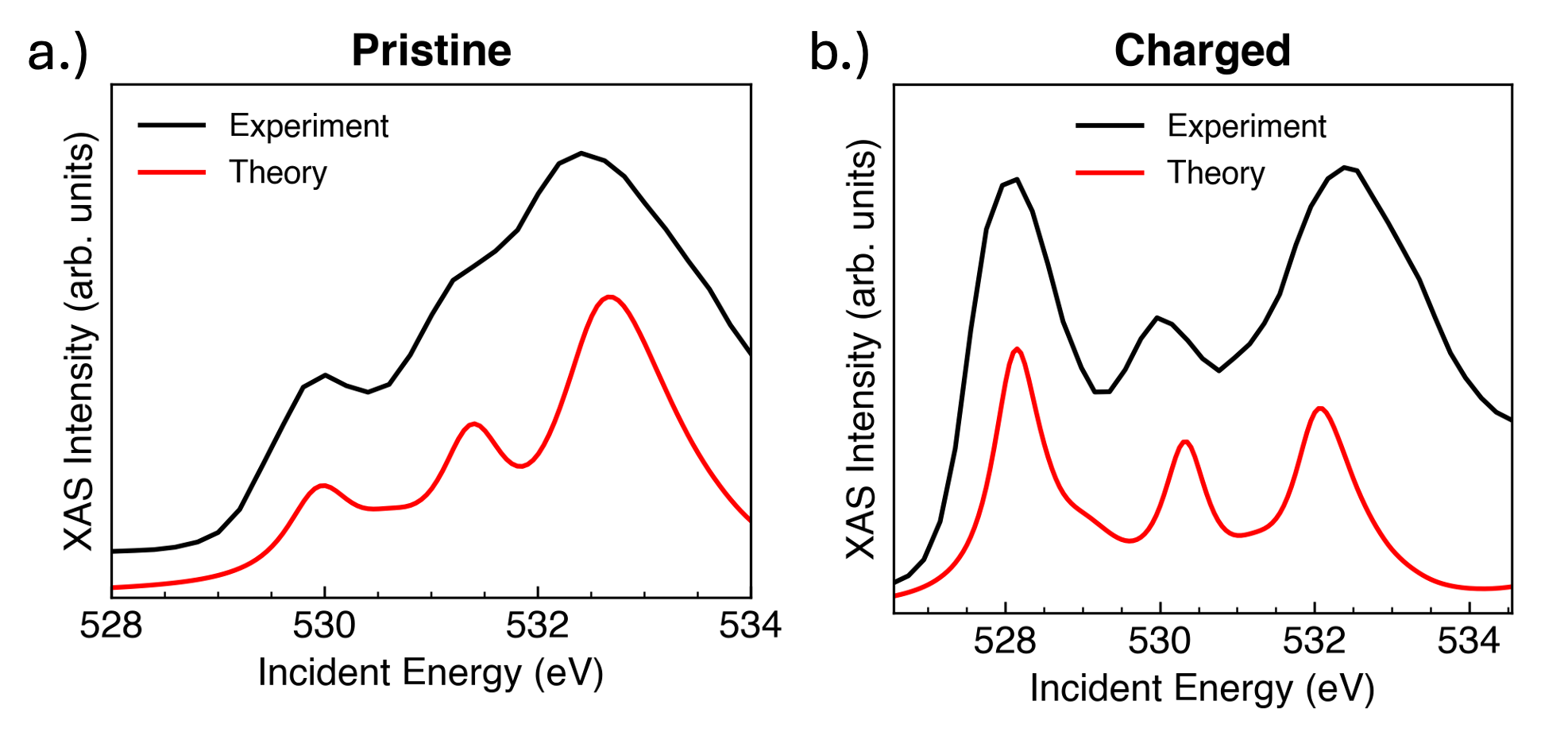}
  \caption[]{\textbf{Theoretical and experimental O $K$-edge XAS spectra} for (a) pristine and (b) charged LFSO. Experimental data comes from Ref.~\cite{ramachandran_formal_2026}. }
  \label{fig:xas}
\end{figure}

\begin{figure}[H]
\centering
  \includegraphics[width=0.94\textwidth]{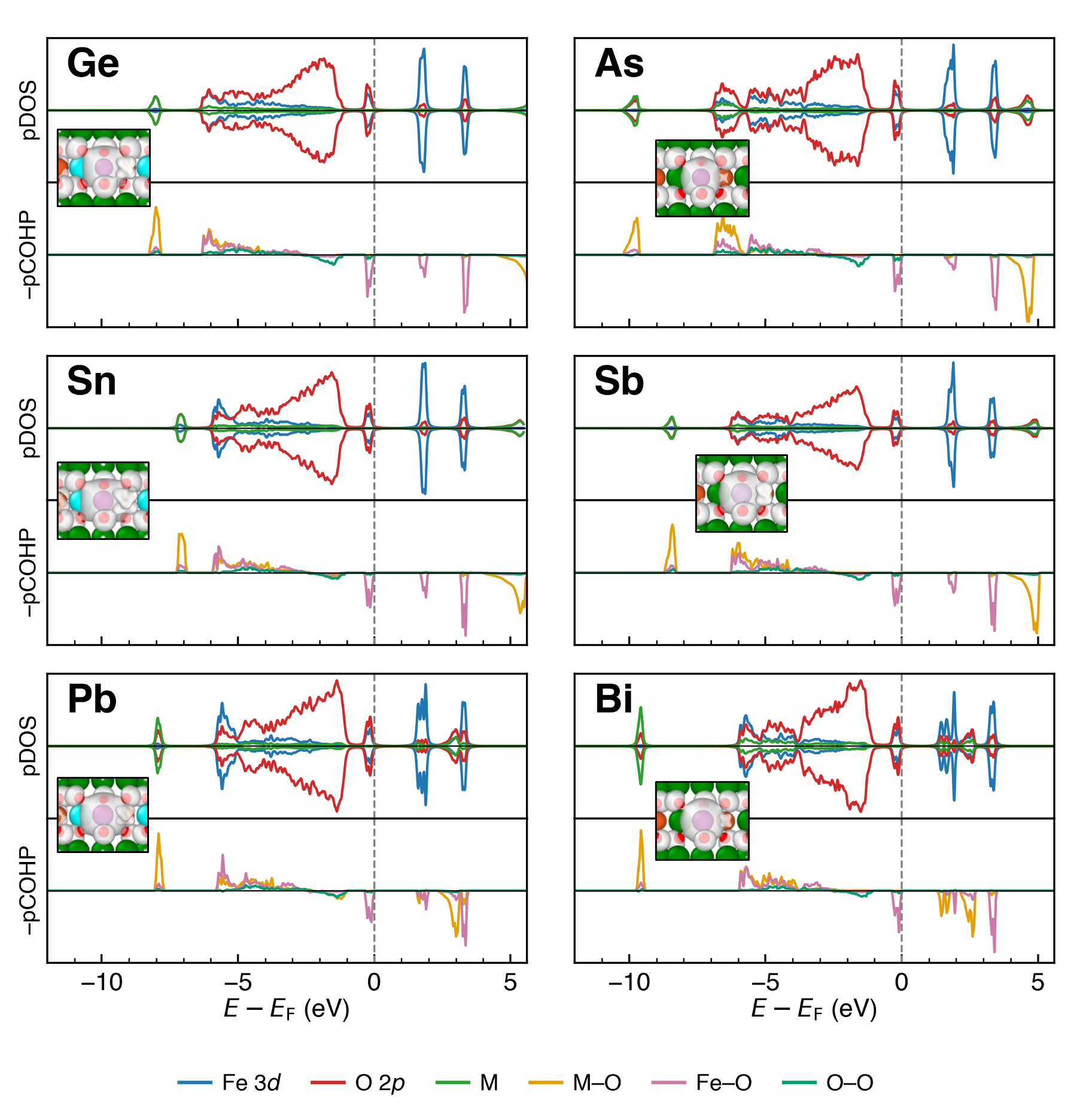}
  \caption[]{\textbf{Full pDOS, pCOHP analysis, and charge density visualization for LFSO and all LFSO-like alternative structures,} labeled according to the element replacing Sb. The covalent metal-oxygen bonding state is present in all structures. }
  \label{fig:pdos_pcohp_6}
\end{figure}

\begin{figure}[H]
\centering
  \includegraphics[width=2.7in]{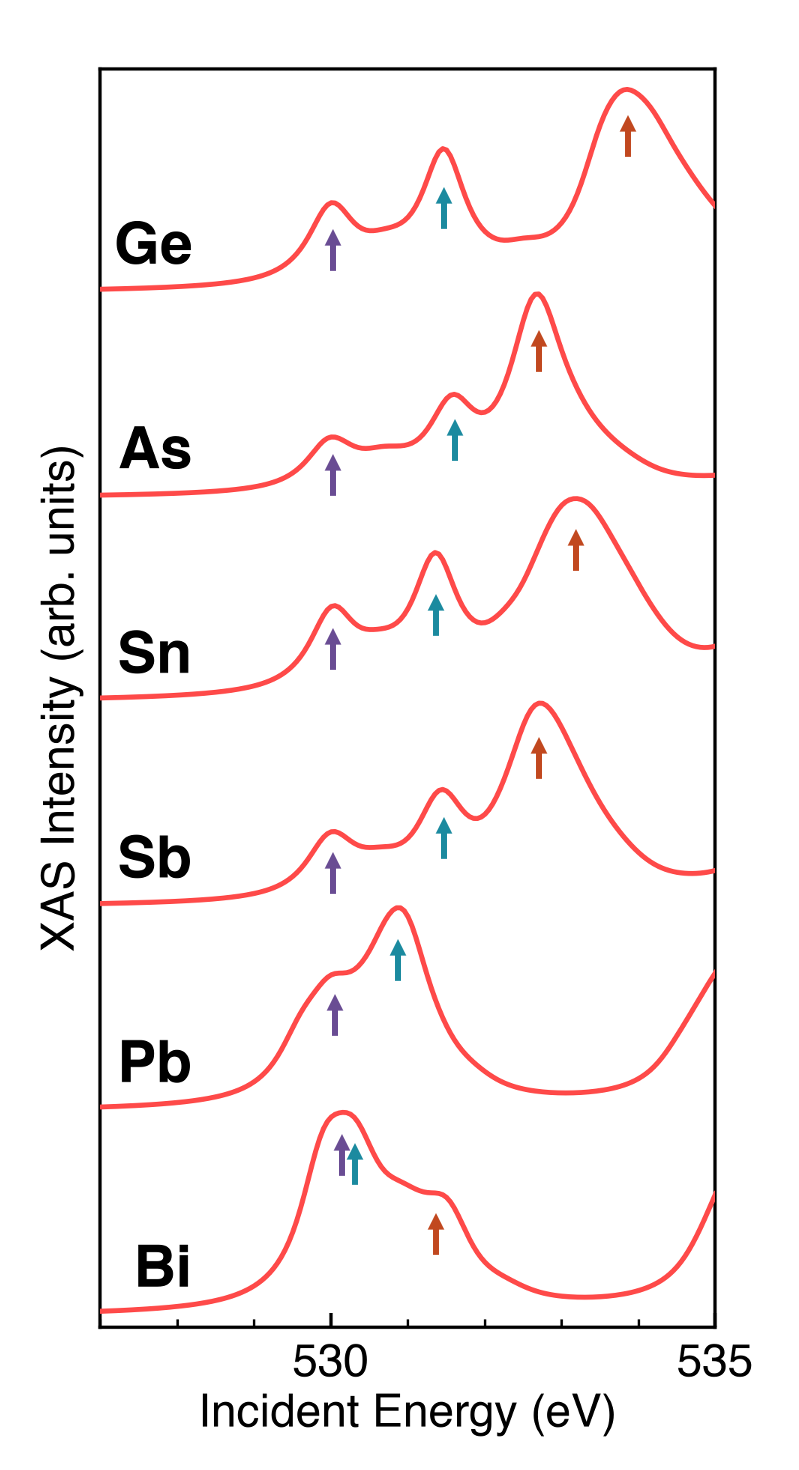}
  \caption[]{\textbf{{\sc ocean}-calculated theoretical XAS spectra for LFSO and all LFSO-like alternative structures,} labeled according to the element replacing Sb. Absolute transition energies from BSE calculations depend arbitrarily on core-hole treatments and DFT references, so we apply a rigid energy shift to align O $K$-edges. The spectrum for Sb is determined by comparison with experiment, and the spectra for all other compounds are shifted to align the first pre-edge peak with Sb. Thus, the comparison focuses on peak positions relative to the first peak. Peaks are highlighted using arrows, with colors corresponding to the RIXS linecuts taken at given incident energies shown in Fig. \ref{fig:rixs_6}.}
  \label{fig:xas_6}
\end{figure}

\begin{figure}[H]
\centering
  \includegraphics[width=\textwidth]{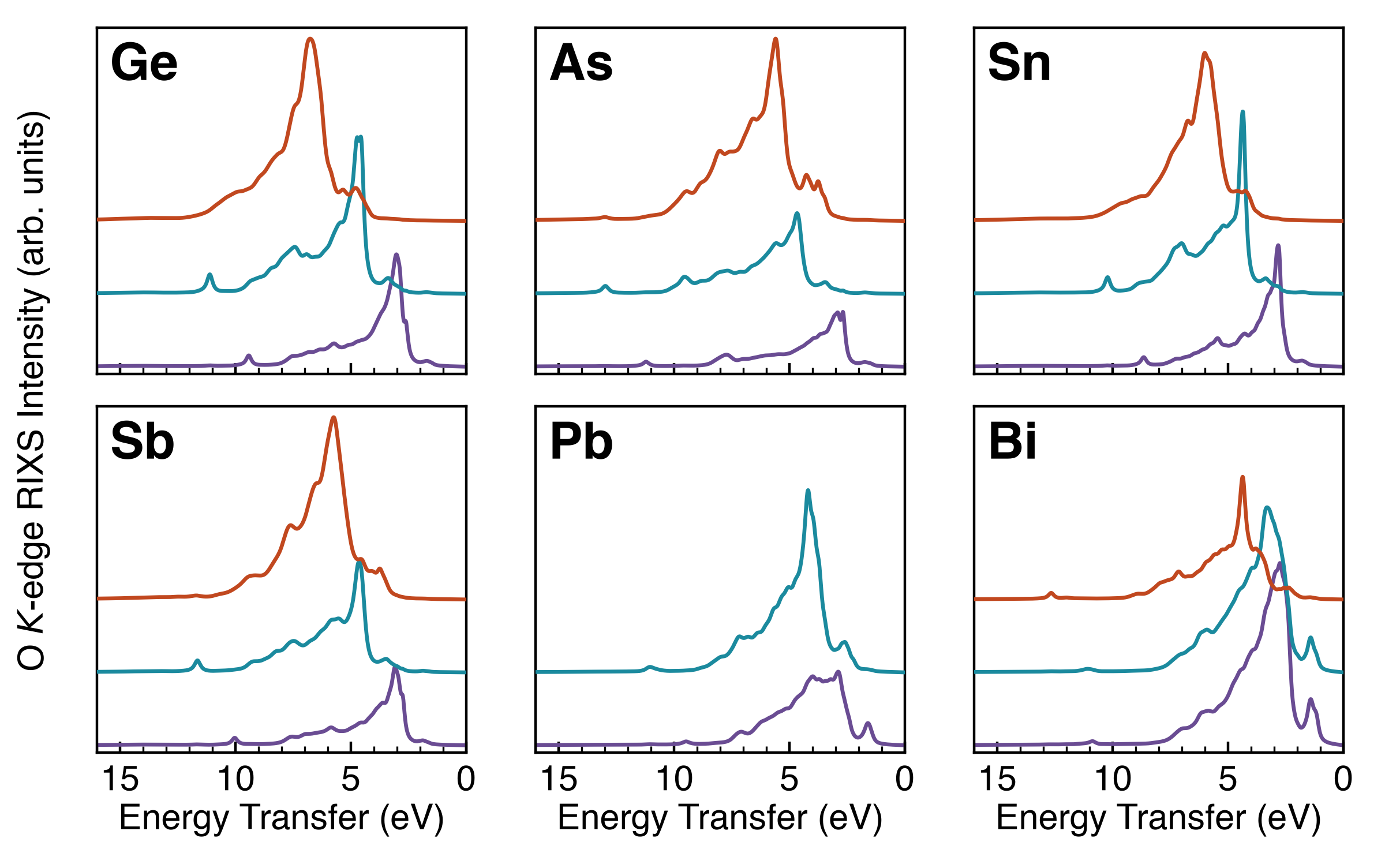}
  \caption[]{\textbf{{\sc ocean}-calculated theoretical RIXS linecuts for LFSO and all LFSO-like alternative structures,} labeled according to the element replacing Sb. Incident energies are chosen according to resonance absorption peaks, and are labeled by color in correspondence with Fig. \ref{fig:xas_6}. The feature corresponding to the covalent $p$-block bonding state can be seen in all panels. The feature is both sharper and stronger for incident energies associated with excitation into Fe-O states, as seen through the pDOS and XAS (Fig. \ref{fig:pdos_pcohp_6} and \ref{fig:xas_6}). }
  \label{fig:rixs_6}
\end{figure}

\begin{figure}[H]
\centering
  \includegraphics[width=5in]{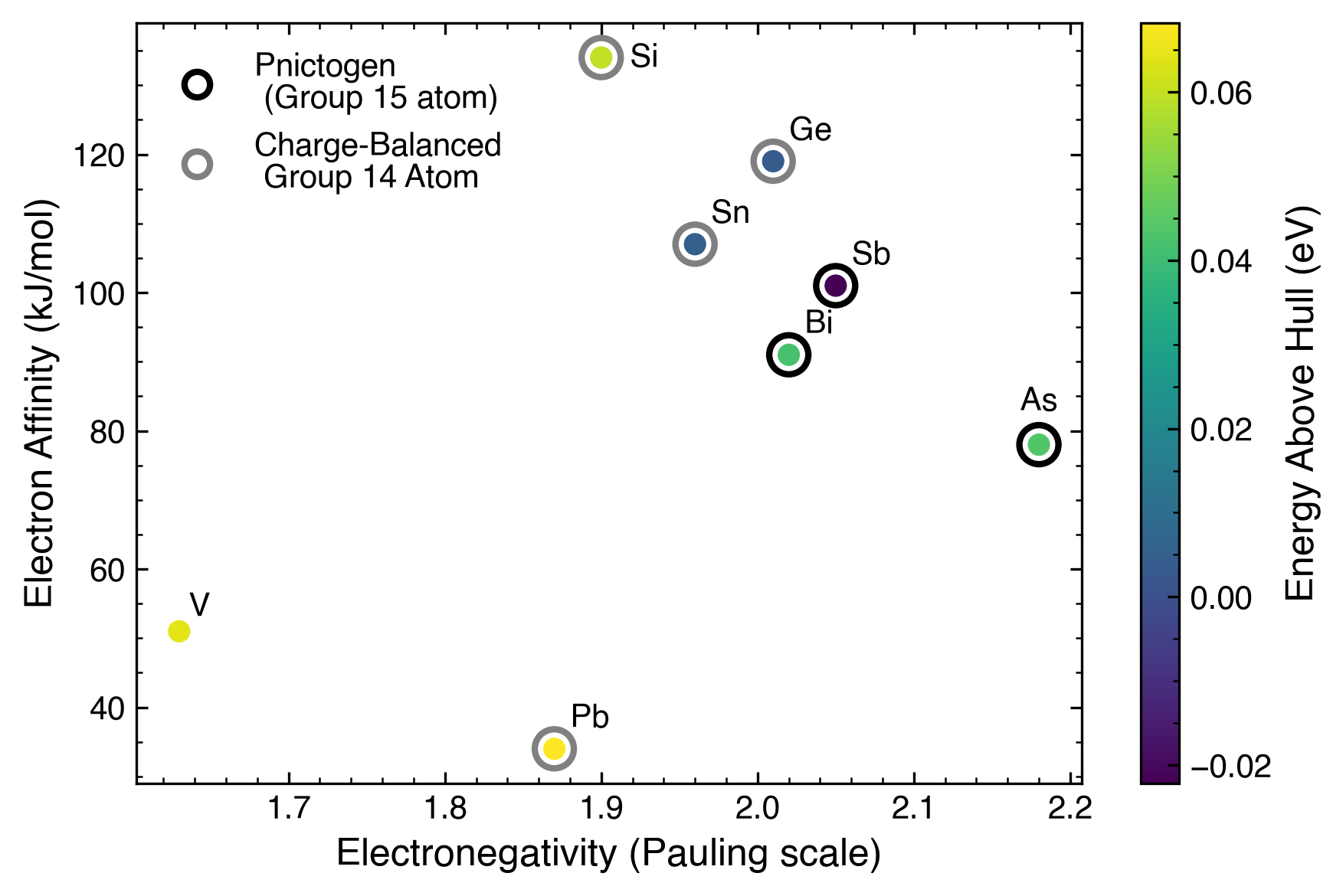}
  \caption[]{\textbf{Thermodynamic convex hull analysis.} Convex hull energy (Table \ref{table:convex_hull}) as a function of electronegativity and electron affinity, two proxies for the strength of covalent bonding between the given element and its coordinated oxygens. All elements are indicated which exhibit the covalent bonding network. }
  \label{fig:thermo_convex_hull}
\end{figure}

\section{Supplementary Tables}

\begin{table}[H]
\centering
\caption{\textbf{Theoretical bond distances and iCOHP values} for selected atomic pairs in LFSO and Sb-replaced candidate structures. More negative iCOHP values indicate interactions which are more bonding. For all structures, there is strong covalency between Sb/Sb-substitute atoms and their coordinated oxygens, and very little covalency between oxygen atoms. Between pristine and charged LFSO, the Sb-O and O-O bond distances and iCOHP values remain relatively constant, even while the Fe-O octahedra undergo significant structural change and negative charge transfer redox.  }
\label{table:bonds}
\begin{tabular}{|c | c | c |c|} 
 \hline
 Structure & Bond Pair & \thead{Average Bond \\ Distance (\AA{})} & \thead{Average iCOHP \\ (eV/ bond)}\\
 \hline
 \multirow{4}{*}{Li$_4$FeSbO$_6$, pristine} 
    & Sb-O & 1.988 & $-5.423$ \\ 
    & Fe-O & 2.029 & $-1.743$ \\ 
    & O-O & 2.861  & $-0.076$ \\
 \hline
 \multirow{4}{*}{Li$_4$FeSbO$_6$, charged} 
    & Sb-O & 1.986 & $-5.436$ \\ 
    & Fe-O & 1.843 & $-3.089$ \\
    & O-O & 2.748 & $-0.084$ \\
 \hline
 \multirow{4}{*}{Li$_4$FeAsO$_6$} 
    & As-O & 1.853 & $-6.205$ \\ 
    & Fe-O & 2.012 & $-1.770$ \\ 
    & O-O & 2.766 & $-0.060$ \\
 \hline
 \multirow{4}{*}{Li$_4$FeBiO$_6$} 
    & Bi-O & 2.112 & $-4.453$ \\
    & Fe-O & 2.038 & $-1.736$ \\
    & O-O &  2.883 & $-0.048$ \\
 \hline
 \multirow{4}{*}{Li$_3$MgFeGeO$_6$} 
    & Ge-O & 1.916 & $-5.606$ \\
    & Fe-O & 2.024 & $-1.756$ \\
    & O-O & 2.806 & $-0.071$\\
 \hline
 \multirow{4}{*}{Li$_3$MgFeSnO$_6$} 
    & Sn-O & 2.069 & $-4.777$ \\
    & Fe-O & 2.043 & $-1.717$ \\
    & O-O &  2.885 & $-0.069$\\
 \hline
 \multirow{4}{*}{Li$_3$MgFePbO$_6$} 
    & Pb-O & 2.172 & $-4.171$ \\
    & Fe-O & 2.049 & $-1.699$ \\
    & O-O & 2.916 & $-0.065$ \\
 \hline
\end{tabular}
\end{table}

\begin{table}[H]
\centering
\caption{\textbf{Bader charge analysis} for pristine and charged LFSO, and ``pristine" and ``charged" Li$_4$FeVO$_6$ (LFVO). LFVO structures were constructed by replacing Sb with V, and geometrically relaxing using {\sc vasp}. Bader charge calculations were performed using the Henkelman code \cite{henkelman_fast_2006}. Between the two structures, Sb plays a greater role in charge compensation than V. }
\label{table:bader}
\begin{tabular}{|l|l|l|}
\hline
Structure & Bader Charge & Change in Charge \\ 
\hline
LFSO, pristine & \begin{tabular}[c]{@{}l@{}}Li: $+0.903$ \\ Fe: $+1.762$ \\ Sb: $+2.733$ \\ O: $-1.351$ \end{tabular} & \multirow{2}{*}{\begin{tabular}[c]{@{}l@{}}$\Delta$Li = $+0.010$ \\ $\Delta$Fe = $+0.088$\\ $\Delta$Sb = $+0.089$\\ $\Delta$O = $+0.268$\end{tabular}} \\ 
\cline{1-2}
LFSO, charged  &  \begin{tabular}[c]{@{}l@{}}Li: $+0.913$ \\ Fe: $+1.850$ \\ Sb: $+2.822$ \\ O: $-1.083$ \end{tabular} & \\
\hline
LFVO, pristine & \begin{tabular}[c]{@{}l@{}}Li: $+0.900$\\ Fe:$+1.786$\\ V: $+2.265$\\O: $-1.275$ \end{tabular}

& \multirow{2}{*}{\begin{tabular}[c]{@{}l@{}}$\Delta$Li = $+0.008$ \\ $\Delta$Fe = $+0.061$\\ $\Delta$V = $+0.048$\\ $\Delta$O = $+0.279$\end{tabular}} \\ 
\cline{1-2}
LFVO, charged  &  \begin{tabular}[c]{@{}l@{}}Li: $+0.908$ \\ Fe: $+1.847$ \\ V: $+2.313$ \\ O: $-0.996$ \end{tabular} & \\
\hline
\end{tabular}
\end{table}

\begin{table}[H]
\centering
\caption{\textbf{Distances from the thermodynamic convex hull,} constructed using all stable entries in each structure's chemical space as given by the Materials Project \cite{jain_commentary_2013}. Negative energy relative to the convex hull indicates a stable compound with respect to decomposition, and structures are considered to be metastable up to $\sim50$ meV/atom above the convex hull.  }
\label{table:convex_hull}
\begin{tabular}{|c | c |} 
 \hline
 Structure & \thead{Energy Relative to Convex Hull \\ (meV/atom)}\\
 \hline
 Li$_4$FeSbO$_6$ & $-22.164$ \\
 Li$_4$FeAsO$_6$ & +43.508 \\
 Li$_4$FeBiO$_6$ & +41.790 \\
 Li$_3$MgFeSiO$_6$ & +59.715 \\
 Li$_3$MgFeGeO$_6$ & +3.750  \\
 Li$_3$MgFeSnO$_6$ & +4.671 \\
 Li$_3$MgFePbO$_6$ & +68.189\\
 Li$_3$CaFeSnO$_6$ & +47.208\\
 Li$_4$FeVO$_6$ & +64.491 \\
 \hline
\end{tabular}
\end{table}

Certain software is identified in this paper in order to specify the experimental procedure adequately. Such identification is not intended to imply recommendation or endorsement of any product or service by NIST, nor is it intended to imply that the materials or equipment identified are necessarily the best available for the purpose.

\end{document}